# RSC Advances

## ARTICLE

# Oil removal from water-oil emulsions using magnetic nanocomposite fibrous mats

S. Barroso-Solares,[a,b] M. G. Zahedi,[a,b] J. Pinto,[a] G. Nanni,[a] D. Fragouli,[a]* and A. Athanassiou[a]*

Herein we present the fabrication of hydrophobic and oleophilic poly(methyl methacrylate)-based nanocomposite fibrous mats with magnetic properties, and their utilization for the oil removal from stable water-oil emulsions. The incorporation of ferromagnetic iron nanoparticles into the polymeric fibers increases the selectivity and oil removal performance of the fibers, as well as introduces magnetic actuation properties to the material. In all the water-oil emulsions used in this work ranging from 5 to 30 v.%, the functional mats can obtain oil absorption efficiencies up to 90%. The oil removal efficiency can reach nearly 100% with just two successive absorption cycles. The high performance achieved makes the presented material a promising candidate for efficient water-oil emulsions separation.

## Introduction

Oily wastewaters produced by petrochemical, pharmaceutical, polymer, metal, cosmetic, textile, and food industries are currently a persistent and common problem, presenting potential environmental and health risks[1]. There are three main categories of oily wastewaters, and different approaches are used to remediate or prevent their adverse effects, usually by separating and removing the oil from the water[2]. The first category is wastewaters where very small amounts of oil are dispersed in the water, and therefore the oil present is classified as *soluble oil*[2]. The main methods employed for the oil removal in this case are deep treatment with adsorbents, nanofiltration or reverse osmosis, and even advanced oxidation technologies. In the second wastewater category, oil is present as *free oil*[3], term employed when significant amount of oil is present as a separated phase. In this case, the remediation typically requires the use of physical techniques such as gravity separation, absorption, and skimming. The last category of oily wastewaters is the *water-oil emulsions*, where one of the phases is dispersed in the other. In this case, the separation of the phases and the removal of the oil is more difficult than in the previous categories, especially when the emulsions are stabilized by particles or surfactants. For this reason many removal technologies have been applied, such as chemical treatment[4,5], biological treatment[6], membrane filtration[1,7,8], etc.

Generally, the physical separation technologies are more convenient approaches due to their minimal environmental impact, feasibility, low operating cost, and effectiveness. Among the water-oil physical separation techniques, the filtration and absorption processes prevail[9,10,11,12]. In both techniques the structural parameters and surface properties of the materials employed are critical for the improvement of the separation performance. In particular, the materials used as filters or absorbents should present a porous structure simultaneously with a selective wettability so as to allow the separation of oil from water[13,14,15,16,17]. For instance, porous materials that are simultaneously hydrophobic (or even better superhydrophobic) and oleophilic (or even better superoleophilic) are not wetted by water whereas the oil is immediately spread onto their surface, resulting in the oil-water separation[1]. A recent progress towards this direction is the development of materials with superhydrophilicity and underwater superoleophobicity, offering an alternative route for oil/water separation[18,19,20].

In general, in the research for functional materials with controlled porosity and selective wettability the ones obtained from polymer fibers are among the most suitable candidates[21]. The electrospinning is a promising technique for the production of such porous materials, allowing a fine adjustment of both the structural parameters and properties not only of the individual polymeric fibers but also of the whole fibrous mat system. Specifically, the possibility to tune the fibers' diameter and to include nanoparticles (NPs) in them can lead to the formation of superhydrophobic fibrous mats[22]. Furthermore, during the fabrication procedure, fibers can be aligned[23], while superhydrophobic membranes can be formed without any permanent supporting substrate[24]. The combination of the tunable wettability, the controlled connectivity, the fine flexibility and the large surface area, offers the possibility to fabricate functional membranes that can be used for the water-oil separation applications.

There are few examples in the literature on the use of selective fibrous materials to separate water-oil emulsions by filtration processes. For instance, Tai et al.[25] successfully employed an electrospun carbon-based nanofibrous membrane as energy efficient and high flux oil-water emulsion separator with an excellent oil permeability at low applied pressure, due to its highly

[a.] Smart Materials, Nanophysics, Istituto Italiano di Tecnologia, via Morego 30, 16163 Genova, Italy.
[b.] Università degli Studi di Genova, via Balbi, 5, 16126, Genova, Italy.

interconnected open pore structure. Following a different approach Si et al.[26] developed nanofiber-assembled cellular aerogels using polyacrylonitrile (PAN) and $SiO_2$ nanofibers as well as $SiO_2$ NPs. These aerogels exhibit optimal properties such as ultralow density, reusability, superhydrophobic-superoleophilic properties, and high pore tortuosity leading to filtration through an effective gravity driven separation of a surfactant-stabilized water-oil emulsion, with an extremely high separation efficiency.

On the contrary, to the best of our knowledge, fibrous mats have not been applied so far as absorbers for the water-oil emulsions separation. Nonetheless, this approach, widely applied for the remediation of wastewaters with free oil, presents significant advantages, such as the possibility to be applied in-situ, without collecting contaminated water for treatment. In the case of the selective absorption of free oil from aqueous media, the design of suitable fibrous functional materials requires once more the surface of the fibers to be superhydrophobic and oleophilic[27]. Wu et al.[28] presented polystyrene (PS) fibers fabricated by electrospinning with a high oil adsorption capacity (g/g, grams of oil per gram of the absorbent material), and discussed the dependence of the oil absorption capacity on the viscosity and surface tension of the oil solvents, as well as on the diameter of the fibers; they found that PS fibers with small diameter and porous surface are able to absorb diesel oil, silicon oil, peanut oil and motor oil with maximum capacities about 7.13, 81.40, 112.30, and 131.63 g/g, respectively. Peng et al.[29] reported that the oil sorption capacity increases with decreasing the electrospun fibers diameter obtained by a blend of PS and PAN, reaching maximum sorption capacities of 194.85, 131.70, 66.75, and 43.38 g/g for pump oil, peanut oil, diesel oil, and gasoline, respectively. Rengasamy et al.[30] proved that polypropylene fiber-assemblies exhibit higher sorption capacity (84.6 g/g) compared to the kapok and milkweed fibers (61.6 g/g and 44.3 g/g, respectively) at a rather constant porosity ~ 0.98. Finally, it was shown that the optimization of the fiber-features leads to outstanding oil absorption capacities. For instance, Lin et al.[31] reported that electrospun nanoporous PS fibers present a motor oil sorption capacity of 113.87 g/g, approximately 2-3 times higher than that of natural sorbents and nonwoven polypropylene fibrous mats.

Herein the potential use of absorption processes for the separation of stable water-oil emulsions is studied, extending thus the use of fibrous absorbents to these particularly persistent and dangerous for the environment and health oily wastewaters. With this aim poly(methyl methacrylate)-based nanocomposite fibers with suitable wetting properties were developed by a simple and versatile single-step electrospinning process, and tested on stable water-oil emulsions.

The presented results show that the functional fibers can adsorb up to 20 times their weight oil from stable water-oil emulsions with 30 v.% oil concentration when the total emulsion mass is 100 times the fibers' weight or more. Under these conditions, the calculated oil removal efficiency reaches 90% for oil contents in water ranging from 5 to 30 v.%.

## Materials and methods

**Materials.** Fibers were produced starting from amorphous poly(methyl methacrylate) (PMMA) (Mw ~120,000 by GPC, ρ=1.17 g/cm$^3$ at 25 °C, Sigma-Aldrich). Iron nanopowder with hydrophobic carbon shell with particle sizes ranging from 5 to 200 nm and average primary particle size ~30-60 nm was purchased from PlasmaChem (Germany), and employed as nanofiller for the development of the nanocomposite fibers. Chloroform ($CHCl_3$, Purity (GC) > 99.80 %, Sigma-Aldrich) was used as received. Mineral oil (ρ=0.84 g/ml at 25 °C, Sigma-Aldrich) and distilled water were mixed to obtain emulsions, using Span80 (Sorbitan monooleate, non-ionic, viscosity 1200-2000 mPa.s at 20°C, HLB value 4.3±1.0, ρ=0.986 g/ml at 25 °C, Sigma-Aldrich) as an emulsion stabilizer.

**Fabrication of the fibers.** The pure PMMA fibers (PFbs) samples were obtained by loading a syringe (diameter of 4.5 mm and 21G needle) with 1 ml of PMMA dissolved in Chloroform (0.5 g/ml). The electrospinning process was performed at room temperature (RT), placing the syringe into a syringe pump (NE-1000, New Era Pump Systems, Inc.), and using a fixed feed rate of 1000 μl/h. A ground target covered with aluminum foil was used as the collector. The metallic tip of the syringe was placed at a distance of 20 cm from the collector, while a voltage of 12 kV was applied between the needle and the collector.

The magnetic nanocomposite fibers with incorporated iron NPs (MFbs) were fabricated following the same procedure. Briefly, the iron NPs were mixed with chloroform (60 mg/ml) and placed in an ultrasonication bath for 2 h. The nanocomposite solution was obtained by dissolving 0.5 g/ml of PMMA in the aforementioned solution (24 h shaking at 800 rpm and RT). The final solution was used for the electrospinning process, using the same parameters as the ones of the pristine PFbs, and obtaining the MFbs with 10.7 wt.% of NPs.

In both cases, the distribution of the obtained fibers mat over the target was not completely homogeneous. Therefore, the prepared fibers were compacted using a hot plate press, kept at 50°C and 98 kPa for 5 min. The absorption experiments were performed using 0.02 g of PFbs or MFbs pressed fibers, shaped in a rectangular piece with dimensions 4x6 cm$^2$ and thickness about 700 μm.

**Preparation of emulsions.** Initially, an appropriate amount of Span80 (0.5 v.%)[32,33] was dissolved into the oil phase and mixed by shaking. This amount was previously determined as the minimum quantity needed to obtain simultaneously stable emulsions and optimum oil absorption (greater amounts of Span80 slightly reduce the absorption capacity, see Supplementary Information, Figure S.1). Next, the aqueous phase was added into the mixture to form the Span80-stabilized emulsions using a high intensity ultrasonication tip (VCX 750, Vibra cell, SONICS) at 40% amplitude for 15 secs at RT. The concentrations of the emulsions are expressed using the volume percentage of the oil (v.%). The oil absorption experiments were carried out using 2 ml water-oil emulsions with the following oil contents of 5-10-20-30 v.%, which correspond to oil masses of 0.084-0.168-0.336-0.504 g, respectively.

**Experimental techniques.** Average diameter and distribution of the fibers were determined using FIJI / ImageJ[34] on micrographs obtained by scanning electron microscopy (SEM, JEOL Model JSM-6490). The fiber samples were coated with gold using a Cressington 208HR sputter coater (Cressington Scientific Instrument Ltd., U.K.) prior to the SEM observations.

Porosity of the fiber mats ($V_f = 1 - \rho_f/\rho_s$), defined as the ratio between the gaseous phase volume and the total volume of the porous sample, was estimated from the ratio between the bulk density of the fibers ($\rho_f$, determined from the weight and geometrical volume of the pressed fiber mats) and the density of the solid PMMA or the nanocomposite ($\rho_s$, 1.17 and about 1.58 g/cm$^3$, respectively)[35].

The mechanical properties of the fibers were measured from tensile tests performed using an Instron 3365 machine, (extension rate: 1 mm min$^{-1}$, and gauge length: 10 mm). Samples were cut into 4x10 mm$^2$ blocks and five specimens for non-pressed and pressed PFbs were tested.

In order to control the stability of the water-oil emulsions a Nikon 139 Eclipse 80i Digital Microscope was employed. To prepare the samples for the microscope analysis, a small amount of the emulsion (one drop) was placed on a microscope glass slide. Then it was covered by a cover slip and finger pressed to make it as thin as possible. Finally the drop size distribution was analyzed using FIJI / ImageJ[34] on the obtained micrographs.

The wetting properties of the fibers were determined measuring the water contact angle (WCA) and oil contact angle (OCA) using a KSVCAM200 (Kruss, Germany) contact angle goniometer. The average and standard deviation contact angle values were determined from two samples of each kind (PFbs and MFbs), taking 5 measurements for each sample using 5 μl drops.

Absorption capacity (C) of the fibers was determined as follows. Three samples of each kind (PFbs or MFbs) were placed into 2 ml of distilled water, oil, or emulsion for 15 minutes. Absorption experiments on emulsions were carried out under shaking, to improve the contact of the fibers with the dispersed phases. The absorption capacity was calculated dividing the amount of the liquid absorbed by the amount of fibers. To this aim, the weight of the fibers before ($w_0$) and after ($w_1$) being in contact with the water, oil, or emulsion was measured, and employed for the calculation of C as reported in Equation 1[11]. In the case of the emulsions, $w_1$ was measured periodically in diverse time intervals, up to 11 days after the absorption tests. In this way, it was possible to ensure the complete water evaporation from the samples (since after 1 week their weight was stabilized, see Supplementary Information, Figure S.2a), in order to identify separately the amount of absorbed oil and water. All experiments were carried out at RT. Moreover, the oil removal efficiency ($E_f$) at these experimental conditions was calculated by dividing the amount of oil absorbed ($w_1 - w_0$) by the initial oil amount ($O_0$) used to prepare the emulsion, and multiplying by 100 (Equation 2)[2]

$$C = \frac{w_1 - w_0}{w_0} \qquad [1]$$

$$E_f = \frac{w_1 - w_0}{O_0} \times 100 \qquad [2]$$

## Results and discussion

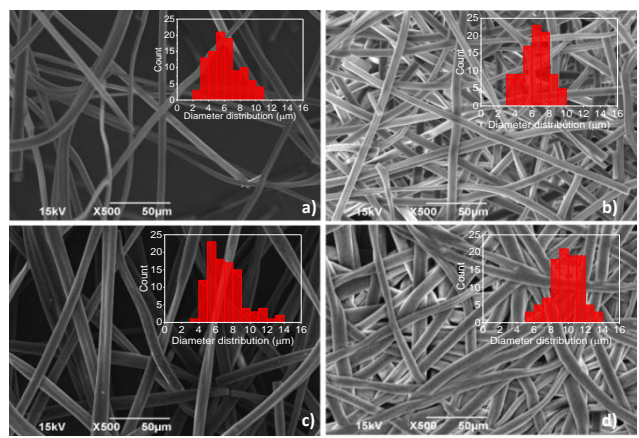

**Figure 1**. *SEM micrographs of: (a) PFbs, (b) pressed PFbs, (c) MFbs, and (d) pressed MFbs.*

In order to study the influence of the magnetic NPs and the impact of the compression process on the fibers' structure and properties, the structural characteristics and surface properties of the PFbs and MFbs mats were investigated. Figure 1 illustrates the morphology, and diameter distribution of the fabricated fibers. The diameter distribution of the PFbs ranges between 2-11 μm, as reported in the histogram (Figure 1a). After pressing, the morphology of the PFbs does not significantly change and their diameter distribution becomes slightly narrower, as shown in Figure 1b. Most importantly, it should be noticed that the pressed fibers retain their individuality and do not show signs of degradation. The addition of the magnetic NPs does not cause any significant change to the diameter distribution of the formed fibers (MFbs), which ranges between 3-14 μm (Figure 1c), whereas for the pressed MFbs the values are slightly higher for the smaller diameters (5-14 μm) due to the pressure exerted. In general, the MFbs present similar aspect, fibers' diameter distribution, and morphology before and after the compression process as in the PFbs case. On the contrary, this compression process had a significant influence on the porosity of the fiber mats; before the compression both the PFbs and MFbs present porosities over 0.90, whereas the pressed fiber mats present porosities below 0.80. Nonetheless, the compression process was necessary in order to improve the fiber mats' structural integrity and the reproducibility of the developed samples, which allow their use as oil absorbers. In fact, during the adsorption tests, the non-pressed fibers are partially damaged or dispersed, hindering their subsequent recovery from the liquid. This is further supported by the study of the mechanical properties of the developed materials. As expected, the pressed fiber mats present a better mechanical performance in terms of Young's modulus compared to the non-pressed ones, showing an enhancement of about 1100% (Supplementary Information, Figure S.3) which can be attributed to the improvement of the entanglement among individual fibers. On the contrary, the non-pressed fiber mats present a poor entanglement among fibers, resulting in a low Young's modulus and a significant dispersion of the fibers in the liquids during the attempts to test their oil absorption performance. For this reason, all the absorption experiments presented below were performed with the pressed fiber mats, assuring the reproducibility and reliability of the proposed method.

The added NPs had a remarkable influence in the wetting properties of the fibrous mats (Figure 2a). The presence of NPs caused the increase of the WCA of the MFbs (WCA = 155±2°) by 25° with respect to the PFbs (WCA = 130±2°). Despite the high WCA, the fibers' surfaces present high water adhesion, which can produce some water absorption only upon immersion of the fibers in water. Therefore, the water absorption capacity of both kinds of fibers was also investigated. After 15 minutes dipped in distilled water, PFbs showed a water absorption capacity of 2 g/g, whereas in the case of MFbs the absorption was 12 % lower, about 1.76 g/g (Figure 2b). Thus, the presence of NPs increases the hydrophobicity of the samples and reduces their water absorption.

In addition, the experiments on the oil wetting properties of the fibrous mats revealed that both kinds of fibers absorb the oil (Figure 2). Nonetheless, the MFbs show a faster oil absorption with respect to the PFbs (an oil droplet of 5 µl is absorbed in ~140 ms by the MFbs and in ~200 ms by the PFbs, Supplementary Information, Figure S.4). After dipping in oil, the MFbs present higher oil absorption capacity compared to the PFbs, with values of 20 g/g and 18 g/g respectively (Figure 2b). Therefore, the introduction of NPs in the polymer fibers (MFbs) is expected to increase the selectivity of the system, reducing the water absorption capacity and at the same time increasing the oil absorption performance. Furthermore, as shown in Figure 3 the MFbs can easily respond to a weak external magnetic field (less than 100 mT) making thus possible their contactless manipulation.

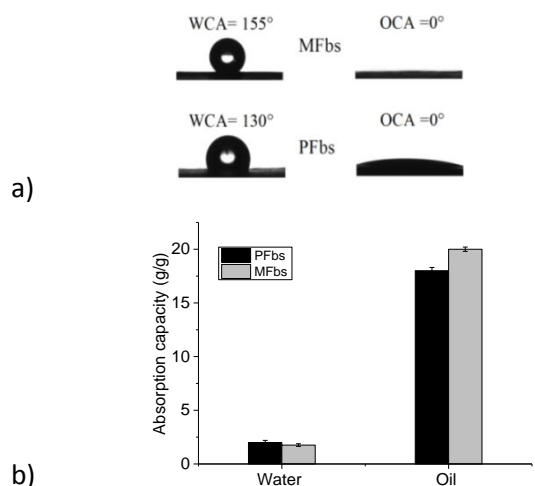

a)
b)

*Figure 2. (a) Water and oil contact angle and (b) water and oil absorption capacity of pressed PFbs and MFbs after 15 min of full immersion.*

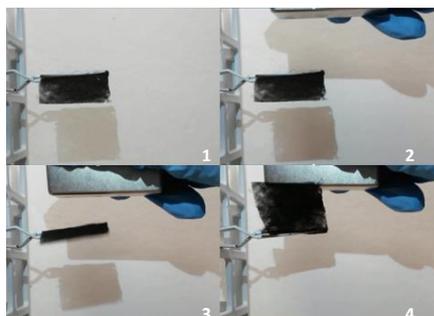

*Figure 3. Image sequence showing the response of MFbs fibers to an external magnetic field generated by a permanent magnet.*

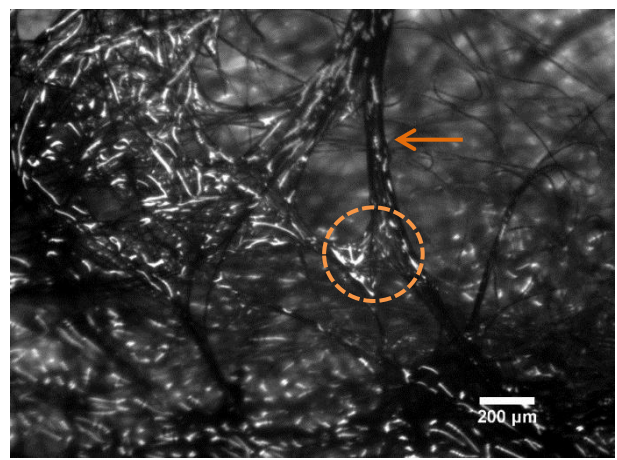

*Figure 4. Optical micrograph of MFbs after their dipping into oil. The circle indicates an area with oil stored in capillary bridges, whereas the arrow shows the surface of one fiber impregnated with the oil.*

The measured oil absorption capacities are lower compared to the ones reported in the literature for other polymeric fiber mats used in absorption of free oil from oil spills (e.g. polypropylene or polystyrene fibers showed a motor oil sorption capacity over 80 g/g)[30][31]. This can be attributed to the porosity of the fiber mats, which in the aforementioned works reaches values up to 0.98; whereas both PFbs and MFbs pressed fibers present porosity below 0.80 and therefore there is less volume available for the oil storage[11]. Indeed, as shown in Figure 4, after the oil absorption tests the oil impregnates completely the fibers' surface, being also stored in capillary bridges between the adjacent fibers[36]. Moreover, the diameters of individual fibers measured do not present significant changes with respect to the diameters before the oil absorption, indicating that there is not swelling of the fibers due to the oil.

Subsequently, the performance of the compacted PFbs and MFbs systems as separators of stable emulsions is presented. Initially, different water-oil emulsions were formed (5-10-20-30 v.% of oil in water) using Span80 as emulsifier. As shown in a representative example of Figure 5a, where is presented the 10 v.% of oil in water emulsion, there are not significant changes in the diameter or number of the formed droplets at different times after the emulsion preparation. By the direct observation of the emulsion micrographs provided in Figure 5b, it can be proved that the diameters of the emulsion droplets show an excellent stability up to 60 minutes. The same behavior was also found for all the concentrations of oil in water prepared using Span80 (see Supplementary Information, Figure S.5). On the contrary, it was found that without Span80 the demulsification starts after 10 min, being possible to observe the phase separation in a macroscopic scale.

The compacted PFbs and MFbs mats were dipped in the stable emulsions, with oil contents from 5 to 30 v.%, and left under stirring for 15 min. Subsequently, after the complete evaporation of the water, the oil absorption capacity was measured and the results are shown in Figure 6. The performance of both types of pressed fibers presents a linear trend, with the oil absorption capacity to increase as the oil concentration in water increases. It was not possible to study this behavior at higher oil concentrations (over 40-50 v.%) due to the partial dispersion of the fibers, independently from their fabrication procedure, in the emulsion during the absorption tests performed with shaking. Therefore, it cannot be experimentally established whether this trend continues or reaches a plateau. However, it can be expected that the oil absorption reaches a plateau for oil contents over 30 v.%, as both types of fibers reached at this point absorption capacities comparable to their maximum oil absorption capacity measured after 15 min in oil (18 g/g for PFbs and 20 g/g for MFbs). As already observed for the isolated liquids, the MFbs present a better performance than the PFbs in the entire range of oil concentrations. In particular, in all cases the increase of the oil absorption capacity for the pressed MFbs is about 2 g of oil per g of fibers compared to the PFbs. This increase is particularly significant at low concentrations of oil (5 v.%), where the capacity of the MFbs becomes 250% higher than that of the PFbS. At the same time the water absorption of the MFbs from stable emulsions with low oil concentrations presents similar values with these of the oil, but for emulsions with high oil concentrations the water absorption is well below the oil absorption (in the case of 30 v.% the absorption capacities are 11.8 g/g and 21.2 g/g for the water and oil, respectively; see Supplementary Information, Figure S.2b).

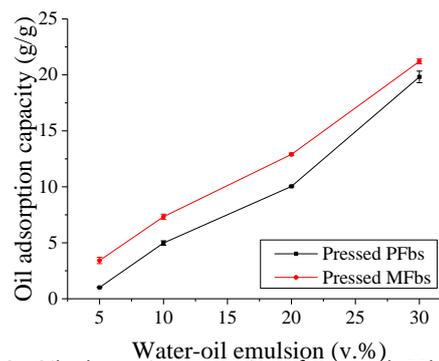

*Figure 6. Oil absorption capacity of pressed PFbs and MFbs in emulsions with oil contents from 5 to 30 v.%. Some of the error bars are within the size of the symbols.*

The higher water absorption capacities for MFbs found after placing the fibers in water-oil emulsions, instead of just in water, can be attributed to the presence of the Span80 surfactant which modifies the surface tension of the water[17]. It should be noticed that this behavior does not hinder the capability of the fibers to absorb oil, as they are able to reach their maximum oil absorption capacity, previously determined by placing them in pure oil. Therefore, it was found that the MFbs are suitable for water-oil emulsions separation, and present better oil absorption performance than PFbs both for free oil or oil in an emulsion.

The suitability of the developed fibers for stable water-oil emulsions separation at different concentrations of oil in water was confirmed by studying the oil removal efficiency $E_f$ obtained when the fibers are employed to treat stable emulsions of a mass up to 100 times higher than the mass of the fibers. This optimal mass ratio was previously determined by studying the oil removal performance from different volumes of emulsions. In the range of emulsions studied, mass ratios lower than 100:1 emulsion:fibers do not further improve the oil removal efficiency and do not allow to reach the maximum oil absorption capacity since the amount of oil is not enough, whereas ratios higher than this decrease the oil removal efficiency and do not increase the maximum oil absorption capacity, as expected (see Supplementary Information, Figure S.6).

As shown in Figure 7 the $E_f$ for all studied emulsions is in general quite high. The MFbs present a rather constant performance, with efficiency values between 80-90%, whereas for the PFbs the efficiency increases with the increase of the oil content in the emulsion, with efficiencies ranging from 20 to 80%. Thus, at low concentrations of oil, e.g. 5 v.%, the MFbs' oil removal efficiency is 3.5 times higher than that PFbs. The MFbs present still a significantly better performance at the 10 and 20 v.% emulsions, being the efficiency respectively 1.5 and 1.3 times higher than the corresponding one for PFbs. In 30 v.% emulsions the efficiency becomes almost identical for the two kinds of fiber mats. This is attributed to the presence of a higher amount of oil in these emulsions, which facilitates the contact between the fibers and the oil, and therefore increases the absorption, without requiring further functionalization. The good performance of the MFbs in emulsions with low oil content makes possible the separation of any surfactant-stabilized emulsion with efficiency near to 100% adopting successive absorption cycles.

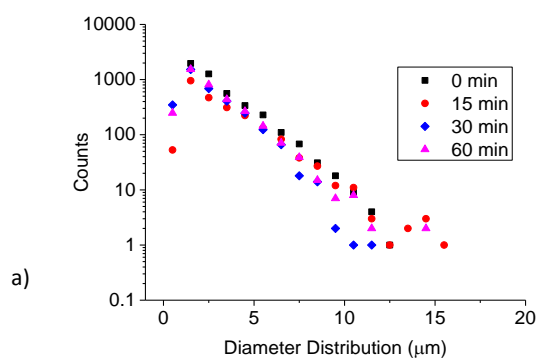

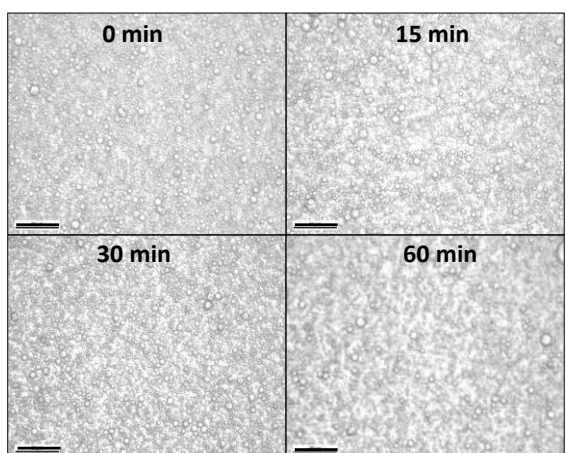

*Figure 5. (a) Drop diameter distribution evolution of 10 v.% water-oil emulsion, (b) sequence of optical micrographs of the emulsion at times from 0 to 60 minutes after the emulsion preparation (scale bar corresponds to 50 µm for all the micrographs).*

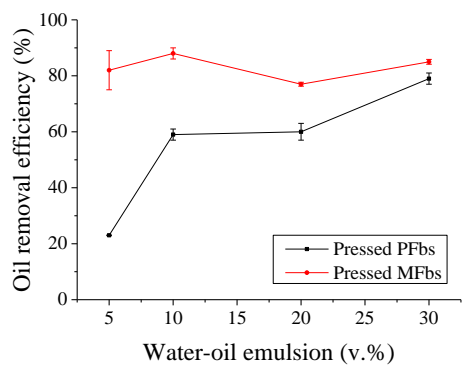

*Figure 7. Oil removal efficiency of pressed PFbs and MFbs in stable emulsions with oil contents from 5 to 30 v.%. Some of the error bars are within the size of the symbols.*

To prove this, it was prepared an emulsion of 30 v.% and subjected to two successive absorption cycles using 0.02 g of MFbs and 2 ml of emulsion for each cycle (following the same aforementioned procedure). After the first absorption cycle it was recovered about 85% of the initial oil, leading to a remaining emulsion with 5 v.% of oil in water, instead of the initial 30 v.%. Then, in the second absorption cycle it was possible to remove about 87% of the oil remaining after the first cycle, leading to a final emulsion with just 0.7 v.% of oil in water. According to that, after two cycles of absorption the overall oil removal efficiency was about 98%. Therefore, it was experimentally demonstrated that the enhancement of the fibers performance introducing the magnetic nanoparticles allows reaching an overall removal oil efficiency near to 100% with just two successive absorption cycles. It should be noticed that this performance is comparable to the widely studied filtration systems based on polymeric membranes, in which separation efficiencies between 96-99% can be also achieved[2,18].

## Conclusions

In this study PMMA nanocomposite fibrous mats were utilized for the oil removal from stable water-oil emulsions. It was demonstrated that the incorporation of ferromagnetic NPs to the polymer matrix offers the desired wettability characteristics by increasing the hydrophobicity and preserving the superoleophilicity of the neat PMMA fibers, as well as induces magnetic actuation properties to the material without modifying the morphology or the diameters of the fibers. Taking advantage of their improved wettability properties, the PMMA nanocomposite fibers were employed for the demulsification process by removing oil from stable water-oil emulsions.

It was found that the presence of nanoparticles improved the performance of the mats reaching absorption capacities up to 20 grams of oil per gram of fibers. Moreover, oil removal efficiencies up to 90% were reached when the fibers are employed to treat stable water-oil emulsions with a mass up to 100 times higher than the fibers' mass, independently from the oil content in water which ranges from 5 to 30 v.%. Finally, taking advantage of these features it was possible to reach oil removal efficiencies about 98% by successive absorption cycles.